\documentclass[intlimits,twoside,a4paper]{article}

\usepackage[cp1251]{inputenc}

\usepackage[eqsecnum]{cmpj3}

\articletype{A part of the Special collection to the 100th anniversary of the birth of Ihor Yukhnovskii}

\issue{2025}{28}{2}{23601}
\doinumber{10.5488/CMP.28.23601}

\title[ Mutual effect of charge- and number-density correlations in ionic liquids]%
{ Mutual effect of charge- and number-density correlations in ionic liquids and concentrated electrolytes}
\author[O. Patsahan, A. Ciach]{O. Patsahan\orcid{0000-0002-5839-3893}\refaddr{label1}\thanks{Corresponding author: \email{oksana@icmp.lviv.ua}.},
	A. Ciach\orcid{0000-0002-5556-401X}\refaddr{label2}\thanks{Corresponding author: \email{aciach@ichf.edu.pl}.}
	}
\addresses{
	\addr{label1} Institute for Condensed Matter Physics of the National
	Academy of Sciences of Ukraine, 1 Svientsitskii Str., 79011 Lviv,
	Ukraine
	\addr{label2} Institute of Physical Chemistry, Polish Academy of Sciences, 01-224 Warszawa, Poland
}


\date{Received January 23, 2025, in final form April 5, 2025}
\begin{document}
	
	\maketitle

\begin{abstract}
Correlation functions in concentrated ionic systems are studied within the mesoscopic theory at the level of the Gaussian approximation. The previously neglected fluctuation contribution to the inverse charge-charge correlation function is taken into account to verify the accuracy of the previous results. We calculate the correlation lengths and the amplitudes and show that the fluctuation contribution does not lead to significant changes of the results. We also derive necessary conditions for the presence of both, the oscillatory and the monotonic decays of the charge-charge correlations that must be satisfied by the  noncoulombic contributions to the  inverse charge-charge correlation function. 
At the level of the Gaussian approximation, these conditions are not satisfied. Extension of the theory beyond the Gaussian approximation is necessary to  verify whether the asymptotic decay of the charge-charge correlations is monotonous or oscillatory, as suggested by the surface force apparatus or by the SAXS experiments, respectively.
%
%
\printkeywords  
%
\end{abstract}

 \section{Introduction}

 Concentrated ionic systems such as water in salt electrolytes or ionic liquids (IL) attract increasing attention in the recent years because they can find various practical applications \cite{Seddon1997,fedorov:14:0,Suo2015,Zhang2016,Watanabe2017,Lei2024}. Classical theories~\cite{debye:23:0}, however, were developed for  dilute electrolytes, and their predictions, although correct for small densities of ions, are completely wrong when the distance between the ions becomes comparable with their diameter. This is because the assumption of point charges is not valid, and oppositely charged neighbors occur with a larger probability than likely charged neighbors. It's already nearly 100 years ago that Kirkwood hypothesized that  the monotonous decay of charge-charge correlations should change to an oscillatory decay for sufficiently large density of ions~\cite{kirkwood:36:0}. The transition from the monotonous to the oscillatory decay is now known as the Kirkwood line~\cite{leote:94:0}. The period and the range of the correlation function are important parameters characterizing the charge ordering that in turn may influence conductivity, capacitance, etc.
 
The question of the structure of concentrated electrolytes and its effect on various physical properties is a big challenge for experiment, simulations and theory. 
Unfortunately, different experimental techniques lead to significantly different results~\cite{Gebbie2013,smith:16:0,lee2017,Hjalmarsson2017,Gaddam2019,Han2018,kumar:22:0,groves:24:0}. Similarly, different simulations and approximate theories lead to the results that disagree with one another \cite{Kjellander2018,Coupette2018,Rotenberg_2018,Adar2019,Coles2020, Cats2021,Outhwaite2021,Zeman2020,Zeman2021,Yang2023,Jger2023,KruckerVelasquez2021,Elliott2024,Dinpajooh2024:2}. Thus, there is no consensus concerning the validity of experimental techniques, simulation methodologies and theoretical assumptions, and the fundamental question of the structure of concentrated electrolytes in various thermodynamic states remains  open.

A lot of discussion was initiated by the measurements of disjoining pressure between crossed mica cylinders confining various electrolytes or IL mixtures~\cite{Gebbie2013,smith:16:0,lee2017}. For sufficiently large $l_\text{B}\rho$, where $l_\text{B}\propto \beta=1/(k_\text{B}T)$ and $\rho$ are the Bjerrum length and the number density of ions, respectively, the decay length $\lambda_s$ of the force satisfied the scaling $\lambda_s/\lambda_\text{D}\sim (a/\lambda_\text{D})^n$ where $n=3$ and $\lambda_\text{D} \propto 1/\sqrt{l_\text{B}\rho}$ is the Debye length. In these experiments, the force decays in an oscillatory way for short distances between the cylinders, and for large distances, a monotonous decay with a very small amplitude was observed. 
Atomic force microscopy (AFM) did not confirm the anomalous underscreening for silica surfaces~\cite{kumar:22:0}. We should note that unlike the charged mica surfaces, the silica surfaces were  weakly charged or neutral. Finally, the authors of recent small-angle X-ray scattering (SAXS) experiments concluded that the asymptotic decay of correlations is oscillatory, and  $n\approx 1.5$ for the considered conditions \cite{dinapajooh:24:0}.

It is  very difficult to obtain the asymptotic decay of charge-charge correlations $G_{cc}(r)$ from simulations, and different results were reported~\cite{Coles2020, Zeman2020,Zeman2021,KruckerVelasquez2021,Elliott2024,Hrtel2023,Dinpajooh2024:2}. Very recently, the oscillatory decay followed by a monotonous decay with a small amplitude was obtained in simulations by H\"{a}rtel, and his result was fitted to a sum of oscillatory and monotonous decays \cite{hartel2024}. Such a sum is a possible solution of the equation derived by Kjellander in his dressed-ion theory \cite{hartel2024}.

On the theoretical side, different approximations were done by different authors.  Kjellander developed the dressed-ion theory, but in practice it is difficult to obtain explicit results from his equations \cite{Kjellander2018,Kjellander2019}. The DFT predictions are consistent with the AFM measurements and yield $n=1.5$ \cite{Yang2023,Jger2023}. With the size of the ions taken into account in calculations of the electrostatic energy, $n=2$ was obtained by Adar et. al. \cite{Adar2019}.  In a  recent article \cite{wang:24:0},  the effect of ion association on the screening behaviour of confined electrolytes was studied using the DFT.
It was found that $n = 1.5$ in the model without association, which is contributed by the charge correlation, and $n = 3$ in the  model with association, which is contributed by the density correlation.
 In our mesoscopic theory, both the size of the ions and the variance of the local charge are taken into account. Our results do not agree with the simple scaling \cite{ciach:23:1}. Rather, the obtained decay length of $G_{cc}(r)$  can be approximated by $\lambda_s/\lambda_\text{D}\sim (a/\lambda_\text{D})^n$ only at short intervals, with $n=1.5$ close to the Kirkwood point, and $n=3$ for $a/\lambda_\text{D}>2.5$. The asymptotic decay of  $G_{cc}(r)$ in our theory is oscillatory.
Since different results were obtained by different methods, the question of the dependence of the type and degree of order on the density of ions and temperature remains open. 

In this work, we reconsider our mesoscopic theory. In our previous studies, the effect of density-density correlations on the charge-charge correlations was neglected, because the corresponding term was expected to be negligible compared to the remaining terms in the expression for the inverse correlation function. In this work we verify whether the assumption that significantly simplified the calculations was justified. We show that the charge-charge correlations are weakly affected when the term neglected in our previous studies is included at the level of the Gaussian approximation. 

We should mention that although our mesoscopic theory was developed independently,  it can be considered as a particular approximation of the exact collective variables (CV)  theory developed by Prof. Yukhnovskii. The CV  method, proposed initially in the 1950s \cite{Zubarev1954,yukhnovskii:58:0} for the description of the classical
charged  particle systems, was developed later for the study of a number of problems  of statistical physics, in particular, for the study of the second order phase transition  \cite{yukhnovskii:80:0,yukhnovskii:87:0}.  
 In \cite{Patsahan2006},  a link between the mesoscopic theory and the CV theory was established for  the case of the
 restricted primitive model of electrolytes with additional short-range interactions. More specifically,  it was shown that after some approximations in the exact microscopic CV action one can arrive at the functional of the grand thermodynamic potential considered in  \cite{ciach:05:0}.
 
 The framework of the mesoscopic theory for electrolytes and IL is briefly summarized in the next section (section~\ref{math}). In section~\ref{math} we also present the equations for the inverse correlation functions in the Gaussian approximation. These equations should be solved self-consistently, but in practice it is a technically difficult task. In section~\ref{assumptions} we present our assumptions and approximations that are valid in a limited range of $\rho l_\text{B}$, but significantly simplify the problem. Explicit expressions for the parameters characterizing the asymptotic decay of the correlation functions are presented in section~\ref{explicit}. The results are shown in section~\ref{results}. In section~\ref{discussion}, we discuss the question of the asymptotic decay of the charge-charge correlations at a more general level.
 In general, the inverse charge-charge correlation function in Fourier representation, $\hat C_{cc}(k)=1/\hat G_{cc}(k)$
 contains a  $k$-dependent contribution  associated with the Coulomb potential, and additional contribution $\beta J_{cc}(k)$ resulting from short-range interactions or from fluctuations.
 We obtain necessary conditions that must be satisfied by $\beta J_{cc}(k)$ in order to obtain  the asymptotic monotonous decay of  $G_{cc}(r)$ for large $\rho l_\text{B}$ in addition to the oscillatory decay with a shorter range. We show that these conditions cannot be satisfied in our theory at the level of the Gaussian approximation. The last section contains our conclusions.

\section{Brief summary of the mesoscopic theory for electrolytes and IL}
\label{math}
The formalism of the mesoscopic theory was developed in a series of works and was applied to different systems with competing interactions in~\cite{ciach:08:1,ciach:11:2,ciach:18:0,patsahan:22:0,ciach:23:1}. In this theory, we consider  mesoscopic densities $\rho_i({\bf r})=6\zeta_i({\bf r})/\piup$, where $\zeta_i({\bf r})$ is the fraction of the mesoscopic volume around ${\bf r}$ covered by the particles of the species $i$.
 The main idea of this theory is to perform the summation over all microscopic states in the expression for the grand potential in two steps. In the first step, we sum over all microscopic states compatible with the given $\rho_i({\bf r})$, and in the next step over all fields $\rho_i({\bf r})$. The first summation leads to the Boltzmann factor with the microscopic Hamiltonian replaced by the grand potential, $\Omega_{co}$, in the presence of the constraints imposed on the microscopic states, i.e., with fixed $\rho_i({\bf r})$. Since the mesoscopic fluctuations are frozen when  $\rho_i({\bf r})$ is fixed, the resulting grand potential can have the mean-field form. In the second step,  the summation over all the mesoscopic densities of $\exp(-\beta \Omega_{co})$ is performed. For the summation of  $\exp(-\beta \Omega_{co})$ over the smooth fields  $\rho_i({\bf r})$, we can apply the field-theoretic methods. In particular, we can make the self-consistent Gaussian approximation. The systematic construction of the mesocopic theory is described in detail in ref.~\cite{ciach:23:1}.

 In the case of ionic systems, we identify the mesoscopic volume with the cube having the edge equal to the ionic diameter $a$. For simplicity, we limit ourselves to the restricted primitive model (RPM) of charged hard spheres with equal diameter $a$ and the same valence of the anions and the cations in a structureless solvent with the dielectric constant $\epsilon$. This model proved to be valid for small size asymmetry with $a=(a_++a_-)/2$ and with specific interactions negligible compared to the Coulomb potential. Due to the symmetry of the RPM, it is convenient to introduce a dimensionless number and charge  density, 
\[
\rho=\rho_++\rho_-, \quad c=\rho_+-\rho_- .
\]
In the self-consistent Gaussian approximation, the grand potential functional of  $\bar c,\bar\rho$ takes, in $k_\text{B}T=1/\beta$ units, the form 
 \begin{equation}
  \label{F}
   \beta \Omega[\{\bar c,\bar\rho\}]=\beta\Omega_{co}[\{\bar c,\bar\rho\}]-\ln \int D\phi\int D\psi \exp\Big(-\beta H_{G}[\{\bar c,\bar\rho\;\phi,\psi\}]\Big),
  \end{equation}
where
 \begin{equation}
    \label{Omco}
    \Omega_{co}[\{\bar c,\bar\rho\}]=
   U_{co}[\bar c]+\int \rd{\bf r} f_h(\bar c({\bf r}),\bar \rho({\bf r}))
   -\mu\int \rd{\bf r}\bar \rho({\bf r}) .
\end{equation}
The first and the second terms in (\ref{Omco}) represent the internal energy 
and the entropy contributions, respectively, in the presence of the constraints $\bar c, \bar\rho$  imposed on the microscopic states, and $\mu$ is the chemical potential of the ions. 
$ f_h$  is the free-energy per unit volume 
of the mixture of hard-spheres with equal sizes in the local-density approximation with
\begin{displaymath}
\label{fh}
\beta f_h=\rho_+\ln \rho_++\rho_-\ln\rho_-+\beta f_{\text{ex}}(\rho),
\end{displaymath}
 and we assume the Carnahan-Starling approximation for $\beta f_{\text{ex}}(\rho)$. 
 If only the Coulomb potential is taken into account, then
the internal energy for a given $c$ is
\begin{equation}
\label{Ucoc}
\beta U_{co}[c]=\frac{l_\text{B}}{2}\int \rd{\bf r}_1\int \rd{\bf r} c({\bf r}_1)\frac{\theta(r-1)}{r}c({\bf r}_1+{\bf r})=\frac{l_\text{B}}{2}\int \frac{\rd{\bf k}}{(2\piup)^3} \frac{4\piup \cos k}{k^2}  \hat c({\bf k})\hat c(-{\bf k}) ,
\end{equation}
where $\hat c({\bf k})$ denotes the function $c$ in Fourier representation, $k=|{\bf k}|$,  and
\begin{displaymath}
    l_\text{B} = \frac{ e^2}{k_\text{B}T\epsilon a}
\end{displaymath}
 is the Bjerrum length in $a$-units. Length and $k$ are in $a$ and $1/a$ units, respectively, in the whole article. In (\ref{Ucoc}), the contributions to the internal energy from overlapping cores of the ions are not included.  

The second term in (\ref{F}) is the contribution from the mesoscopic fluctuations 
 $\phi({\bf r})=c({\bf r})-\bar c$ and $\psi({\bf r})=\rho({\bf r})-\bar \rho$ of $c$ and $\rho$, respectively, and in the self-consistent Gaussian approximation 
  \begin{eqnarray*} 
  \beta H_G[\bar c,\bar\rho,\phi,\psi]=\frac{1}{2}\int \rd{\bf r}_1\int \rd{\bf r}_2
\Big(  \phi({\bf r}_1)C_{cc}(r) \phi({\bf r}_2)+ \psi({\bf r}_1)C_{\rho\rho}(r) \psi({\bf r}_2)+2 \phi({\bf r}_1)C_{c\rho}(r)\psi({\bf r}_2) \Big),
  \end{eqnarray*}
with $r=|{\bf r}_1-{\bf r}_2|$,
 \begin{displaymath}
 C_{cc}(r)=\frac{\delta^2\beta\Omega}{\delta\bar c({\bf r}_1)\delta\bar c({\bf r}_2)}
\end{displaymath}
and with $C_{\rho\rho}(r)$ and $C_{c\rho}(r)$ given by analogous expressions.
Note that with the Boltzmann factor proportional to $\exp(-\beta H_G)$, the correlation functions $\langle \phi({\bf r}_1)\phi({\bf r}_2)\rangle=G_{cc}(|{\bf r}_1-{\bf r}_2|)$ and  $G_{\rho\rho}(|{\bf r}_1-{\bf r}_2|)=\langle \psi({\bf r}_1)\psi({\bf r}_2)\rangle$ in Fourier representation are $\hat G_{cc}(k)=1/\hat C_{cc}(k)$ and $\hat G_{\rho\rho}(k)=1/\hat C_{\rho\rho}(k)$, respectively, since in equilibrium $c=0$ and $C_{c\rho}=0$.

Note that  there are two contributions to $C_{cc}$ from the two terms in $\Omega$ [see (\ref{F})]. In the fluctuation contribution to  $C_{cc}$, we have functional derivatives of the functionals $C_{cc}$,  $C_{\rho\rho}$ and $C_{c\rho}$, multiplied by the correlations of the fluctuations~\cite{ciach:23:1}. 
 We  neglect the fluctuation contribution to the higher-order functional derivatives of $\beta\Omega$ (i.e., the functional derivatives of $C_{\alpha\beta}$) and for  $n+m>2$ assume
\begin{eqnarray*} 
 \frac{\delta^{n+m}(\beta \Omega)}{\delta c({\bf r}_1)\ldots\delta c({\bf r}_n)\delta \rho({\bf r}_{n+1})\ldots\delta \rho({\bf r}_{n+m})}&\approx&
 \frac{\delta^{n+m}(\beta \Omega_{co})}{\delta c({\bf r}_1)\ldots\delta c({\bf r}_n)\delta \rho({\bf r}_{n+1})\ldots\delta \rho({\bf r}_{n+m})}
\\
\nonumber
&=& A_{m,n}(c,\rho)\delta({\bf r}_1-{\bf r}_2)\ldots\delta({\bf r}_{n+m-1}-{\bf r}_{m+n})
\end{eqnarray*}
where
\begin{displaymath}
\label{Amn}
 A_{m,n}(c,\rho)=\frac{\partial^{n+m}(\beta f_h)}{\partial^n c\partial^m \rho}.
\end{displaymath}
This approximation is analogous to the self-consistent one-loop approximation in the field-theoretic approach. 

Note that the charge waves with the wavelength $2\piup/k$  are energetically favored for $k>\piup/2$ (see (\ref{Ucoc})), because oppositely charged close neighbors give a negative contribution to the energy, and the variance of the local charge,
 \begin{equation}
 \label{variance}
 \langle\phi^2\rangle= \int \frac{\rd{\bf k}}{(2\piup)^{3}}\hat C_{cc}^{-1}(k)=\int_0^{\piup} \frac{ \rd k }{(2\piup)^{2}}\frac{ 2k^2}{\hat C_{cc}(k)}
 \end{equation}
should be taken into account. In the mesoscopic theory, the cutoff $\Lambda=2\piup/2$ is defined by the shortest wavelength $2a$ of the charge wave.
Local fluctuations of $\rho$ are not energetically favored since the energy does not depend on $\rho$ [see (\ref{Ucoc})], and large local density fluctuations are not expected. Thus,  $\langle\psi^2\rangle$ can be neglected. With the above assumptions, the equations for the inverse correlation functions in the self-consistent Gaussian approximation take the forms~\cite{ciach:21:0}
\begin{equation}
\label{Ccc3}
 \hat C_{cc}(k)\approx  \frac{4\piup l_\text{B}\cos k}{k^2} +
 \frac{1}{\rho}+\frac{1}{\rho^3}\langle\phi^2\rangle +\beta \hat J_{cc}(k)
\end{equation}
and
\begin{equation}  
\label{Crr3}
 \hat C_{\rho\rho}(k)\approx
 A_{2,0}+\frac{1}{\rho^3}\langle\phi^2\rangle  +\beta \hat V_{fl}(k),
\end{equation}
with  $\langle\phi^2\rangle$ given in (\ref{variance}),
\begin{equation}
\label{Jcc}
 \beta \hat J_{cc}(k)=  -\frac{1}{\rho^4}\int \rd{\bf r}\, \re^{\ri{\bf k}\cdot{\bf r}} G_{cc}(r)G_{\rho\rho}(r)
\end{equation}
and
\begin{equation}
\label{Vfl}
\beta\hat V_{fl}(k)=-\frac{A_{1,2}^2}{2}\int \rd{\bf r}\,\re^{\ri{\bf k}\cdot{\bf r}}G_{cc}(r)^2-\frac{A_{3,0}^2}{2}\int \rd{\bf r}\,\re^{\ri{\bf k}\cdot{\bf r}}G_{\rho\rho}(r)^2.
\end{equation}

The last term in (\ref{Ccc3}) was neglected in our previous studies, because it is proportional to the product of the amplitudes of the charge and density correlations. The fluctuation-induced effective interactions $ V_{fl}$ in (\ref{Crr3})  are proportional to the square of the amplitude of the charge-charge correlations. Thus, the last term in  (\ref{Ccc3}) is expected to be of a third order in the amplitude of the charge-charge correlations. For this reason we assumed that this term is much smaller than $\langle\phi^2\rangle/\rho^3$ and can be neglected. The integral, however, is proportional to the decay length of the correlations, and if this length is large, the neglected term may in principle become relevant. Moreover, the neglected term depends on $k$ and therefore may change the period of the charge oscillations.

Note that when the last term in  (\ref{Ccc3}) is neglected, the charge fluctuations influence the number-density correlations, but the charge-charge correlations do not depend on $C_{\rho\rho}$. This makes the problem of self-consistent solution of the two equations much simpler. With the last term in  (\ref{Ccc3}) included, the mutual effect of the charge and density correlations is taken into account. In this work we determine this mutual effect of the correlations in the case of large $\rho l_\text{B}$, where $\hat C_{cc}(k)$ has a deep minimum for $k_0>0$.
 \section{Assumptions and approximations}
\label{assumptions}
We assume that $\beta\hat J_{cc}$ is a small correction, therefore it only weakly influences  the pronounced maximum of $1/\hat C_{cc}(k)$ that we observed for large $l_\text{B}\rho$ in the absence of the last term in (\ref{Ccc3}). 
 For large values of $l_\text{B}\rho$, we can  expand  $\hat C_{cc}(k)$ in a Taylor series about the deep minimum at $k=k_0$, and truncate the expansion.  Since  $\hat C_{cc}(k)$ is an even function of $k$, we can Taylor expand $\hat C_{cc}(k)$ in $k^2$, and make the approximation
  \begin{equation}
\label{Caa}
\int_0^{\piup} \frac{ \rd k}{(2\piup)^2} \frac{2k^2}{\hat C_{cc}(k)}\approx \int_0^{\infty} \frac{ \rd k}{(2\piup)^2} \frac{2k^2}{\hat C_{a}(k)},
\end{equation}
with 
\begin{equation}
\label{Ca}
 \hat C_{cc}(k)\approx   \hat C_{a}(k)=\hat C_{cc}(k_0)+\beta v (k^2-k_0^2)^2 ,
\end{equation}
that is valid for $k\approx k_0$. 
Formally, the above equations are the same as in the absence of the last term in (\ref{Ccc3}). When $\beta\hat J_{cc}$ is taken into account in (\ref{Ccc3}), however,  the parameters $k_0, \hat C_{cc}(k_0)$ and $\beta v$ take on the values that depend on the density-density correlation function [see (\ref{Jcc})] that in turn depends on $\hat C_{cc}(k)$.  
We assume that in the presence of a small correction in (\ref{Ccc3}),  we still have  $1/\hat C_a(k_0)\gg 1$ for large $l_\text{B}\rho$ and  the main contribution to the integral in (\ref{Caa}) comes from the neighborhood of $k_0$. Due to this and since 
 $1/\hat C_a(k)$ decays sufficiently fast for increasing $k>\piup$, the  contribution from $k\in(\piup,\infty)$ to the integral on the right-hand side of (\ref{Caa}) is negligible~\cite{ciach:23:1} and the approximation is justified. We stress that the approximation (\ref{Caa}) is not valid when the maximum of $1/\hat C_{cc}(k)$ is not high and narrow, i.e., when $l_\text{B}\rho$ is not sufficiently large~\cite{ciach:23:1}.
 
  The parameters $k_0,\hat C_{cc}(k_0) $ and $\beta v$  can be obtained from the value of $\hat C_{cc}(k)$ at the minimum given by 
\begin{displaymath}
\left.\frac{\rd\hat C_{cc}(k)}{\rd k}\right|_{k=k_0}=0,
\end{displaymath}
 and from the second derivative at $k_0$. 
The approximation (\ref{Ca}) leads to the correlation function in the real space of the form 
\begin{equation}
\label{Ans1}
G_{cc}(r)\approx G_a(r)=\frac{A_{c}\sin(\alpha_1 r)\re^{-\alpha_0 r}}{r},
\end{equation}
where the parameters $A_c, \alpha_0,\alpha_1$ can be obtained by equating  equation~(\ref{Ca}) with
the inverse Fourier transform of (\ref{Ans1}),  
\begin{equation}
\label{FT}
\hat C_{a}(k)=\hat G_{a}(k)^{-1}=\frac{ 4\alpha_0^2\alpha_1^2+(k^2-\alpha_1^2+\alpha_0^2)^2}{8\piup\alpha_0\alpha_1 A_c}.
\end{equation}
 As the parameters in (\ref{Ca}) are determined by the first and the second derivative of $\hat C_{cc}(k)$ given in (\ref{Ccc3}) and by its value at the minimum, we have 3 equations for $A_c,\alpha_0,\alpha_1$.  Moreover, for the approximation~(\ref{Ans1}) we have $\langle \phi^2\rangle\approx G_a(0)=A_c\alpha_1$.
Note, however, that  $\beta\hat  J_{cc}(k)$ [see (\ref{Jcc})] depends on both, the charge-charge and the density-density correlation functions. Therefore, the equations relating $A_c, \alpha_0,\alpha_1$ with $\hat C_{cc}(k)$ depend in addition on the density-density correlations. 

Let us focus on the density-density correlations [see (\ref{Crr3})]. As noted in \cite{ciach:21:0,ciach:2023:2}, the fluctuation-induced effective interactions $\beta\hat V_{fl}(k)$ take on a negative minimum at $k=0$, and the expansion
\begin{equation}
\label{Vflexp}
\beta\hat V_{fl}(k)\approx \beta\hat V_{fl}(0)+\beta V_2k^2 +O(k^4)
\end{equation}
together with (\ref{Crr3}) lead to the asymptotic decay of the density-density correlations of the form
\begin{equation}
\label{Ans2}
G_{\rho\rho}(r)=\frac{A_{\rho} \re^{- r/\xi}}{r},
\end{equation}
with the corresponding Fourier transform
\begin{equation}
\label{Te}
\hat G_{\rho\rho}(k)^{-1}=\hat C_{\rho\rho}(k)=\frac{1}{4\piup A_{\rho}}\Bigg(\frac{1}{\xi^2}+k^2\Bigg) +O(k^4).
\end{equation} 
The parameters $A_{\rho}$ and $\xi$ can be obtained by equating (\ref{Te}) with (\ref{Crr3}) where $\beta\hat V_{fl}(k)$ should be approximated by (\ref{Vflexp}).

In principle, we have a mathematically trivial problem of solving 5 equations for 5 unknowns $A_c,A_{\rho}, \alpha_0,\alpha_1,\xi$.
In practice, the resulting  expressions  are not so simple. We present the explicit forms of the equations in the next section.

\section{Explicit forms of the self-consistent equations for the correlation functions}
\label{explicit}
Let us first present the explicit expressions for   $\beta\hat J_{cc}(k), \beta\hat V_{fl}(0)$ and $\beta V_2$.  When $G_{cc}(r)$ and $G_{\rho\rho}(r)$ are approximated by (\ref{Ans1}) and (\ref{Ans2}), respectively,  we obtain from (\ref{Jcc})

\begin{eqnarray*}
\beta \hat J_{cc}(k)=-\frac{4\piup A_cA_{\rho}}{\rho^4k}\int_0^{\infty} \rd r \frac{\sin(kr)\sin(\alpha_1 r)\re^{-r/\xi_o}}{r}=-\frac{\piup A_cA_{\rho}}{\rho^4}{\cal F}(k),
\end{eqnarray*}
where
\begin{displaymath}
{\cal F}(k)=\frac{1}{k}\ln\Bigg[
\frac{(\alpha_0+\xi^{-1})^{2}+(k+\alpha_1)^2}{(\alpha_0+\xi^{-1})^{2}+(k-\alpha_1)^2}\Bigg].
\end{displaymath}
This function depends on the characteristic lengths, and does not depend on the amplitudes. For clarity, we do not display all its arguments.

From (\ref{Vfl}), (\ref{Ans1}) and (\ref{Ans2})  we obtain the explicit expressions for the fluctuation-induced effective density-density interactions in the Taylor expansion (\ref{Vflexp}),
\begin{equation}
\label{bVfl}
\beta\hat V_{fl}(0)= -\frac{\piup A_{c}^2}{2\rho^4}
  \frac{\alpha_1^2}{\alpha_0(\alpha_0^2+\alpha_1^2)}-\piup A_{3,0}^2A_{\rho}^2\xi
\end{equation} 
and
\begin{equation}
\label{bV2}
\beta V_2 =\frac{\piup}{3\rho^4}\frac{ A_{c}^2}{8}\Bigg(
  \frac{1}{\alpha_0^3}+\frac{\alpha_0(3\alpha_1^2-\alpha_0^2)}{(\alpha_0^2+\alpha_1^2)^3}
  \Bigg)+ \frac{2\piup A_{3,0}^2A_{\rho}^2}{3}\Bigg(\frac{\xi}{2}\Bigg)^3.
\end{equation}
These parameters depend on all the lengths and the amplitudes.
The above forms will be used in our equations for the parameters of the correlation functions.

To simplify the notation, we introduce 
\begin{displaymath}
B_n(k_0)=\left.\frac{\rd^n (\cos (k)/k^2)}{\rd k^n}\right|_{k=k_0}
\end{displaymath}
and
\begin{displaymath}
F_n(k_0)=\left.\frac{\rd^n{\cal F}(k)}{\rd k^n}\right|_{k=k_0}
\end{displaymath}
for $n=0,1,2$.

The condition for the minimum of $\hat C_{cc}(k)$ takes the form
\begin{equation}
\label{min}
4l_\text{B} B_1(k_0)=\frac{A_cA_{\rho}}{\rho^4}F_1(k_0).
\end{equation}
By equating (\ref{FT}) and (\ref{Ca}) and using the second derivative of $\hat C_{cc}(k)$ we obtain
\begin{equation}
\label{sec}
\hat C_{cc}(k_0)^{''}=4\piup l_\text{B}B_2(k_0)-\frac{\piup A_cA_{\rho} F_2(k_0)}{\rho^4}=\frac{(\alpha_1^2-\alpha_0^2)}{\piup \alpha_1\alpha_{0}A_c}.
\end{equation}
Next, we introduce 
\begin{equation}
\label{G}
\frac{A_{c}A_{\rho}}{\rho^4}=G(k_0)=\frac{4l_{B}B_1(k_0)}{F_1(k_0)},
\end{equation}
where (\ref{min}) was used, and from (\ref{sec}) we obtain the expression for the amplitude in terms of the lengths
\begin{equation}
\label{Ac1}
A_c= \frac{\alpha_1^2-\alpha_0^2}{\piup^2 \alpha_1\alpha_0(4l_\text{B}B_2(k_0)-G(k_0)F_2(k_0))}.
\end{equation}
For $A_{\rho}$ we obtain from (\ref{G})
\begin{equation}
\label{Ar1}
A_{\rho}=\frac{\rho^4 G(k_0)}{A_c},
\end{equation}
where $A_c$ is given in (\ref{Ac1}) as a function of $\alpha_0,\alpha_1,\xi$ and $k_0$.
The remaining equations concerning $\hat C_{cc}(k)$ obtained by equating (\ref{FT}) and (\ref{Ca}) and using (\ref{G}) are
\begin{equation}
\label{k0}
k_0^2=\alpha_1^2-\alpha_0^2,
\end{equation}
\begin{equation}
\label{e1}
\hat C_{cc}(k_0)=4\piup l_{B}B_0(k_0)+\frac{1}{\rho}+\frac{A_c\alpha_1}{\rho^3}-\piup G(k_0)F_0(k_0)=\frac{\alpha_0\alpha_1}{2\piup A_c}.
\end{equation}

Finally, by equating (\ref{Crr3}) with (\ref{Te}) and using (\ref{Vflexp}) we obtain the last two equations
 \begin{equation}
 \frac{1}{4\piup A_{\rho}}=\beta V_2
\end{equation} 
and
\begin{equation}
\label{cr2}
\frac{1}{4\piup A_{\rho}\xi^2}=A_{2,0}+\frac{A_c\alpha_1}{\rho^3}+\beta \hat V_{fl}(0),
\end{equation}
where (\ref{bV2}) and (\ref{bVfl}) should be used.

The parameters in the approximate forms of the correlation functions, (\ref{Ans1}) and (\ref{Ans2}) are solutions of the set of equations (\ref{Ac1})--(\ref{cr2}). Since the amplitudes and $k_0$ are given in (\ref{Ac1})--(\ref{k0}), the 3 equations  for $\alpha_0,\alpha_1,\xi$, (\ref{e1})--(\ref{cr2}), should be solved numerically. 
\section{Results}
\label{results}

\begin{figure}[!t]
	\centering
	\includegraphics[scale=0.3]{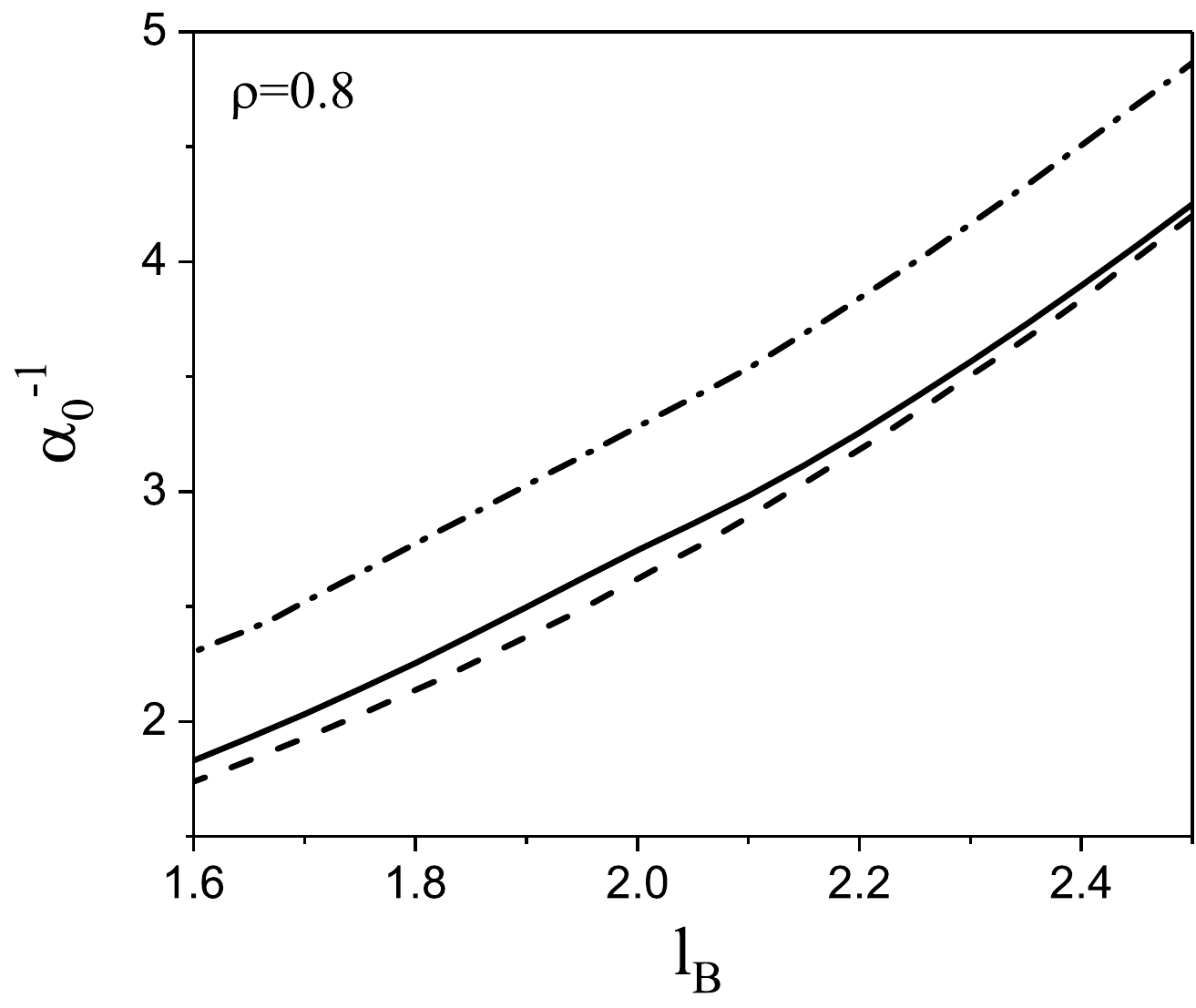}
	\caption{The correlation length  of the charge--charge correlations $\alpha_0^{-1}$ as a function of  the Bjerrum length $l_\text{B}$  for the fixed dimensionless density of ions $\rho=0.8$ obtained with a solution of the equations~(\ref{e1})--(\ref{cr2}) (solid line). The results obtained  without the last term in (\ref{Ccc3})  are shown for comparison: from the approximate analytical theory with the approximation (\ref{Caa}) (dashed line) and with a numerical solution (dash-dotted line)~\cite{ciach:23:1} .}
	\label{fig:alpha0^-1} 
\end{figure}

\begin{figure}[!t]
	\centering
	\includegraphics[scale=0.25]{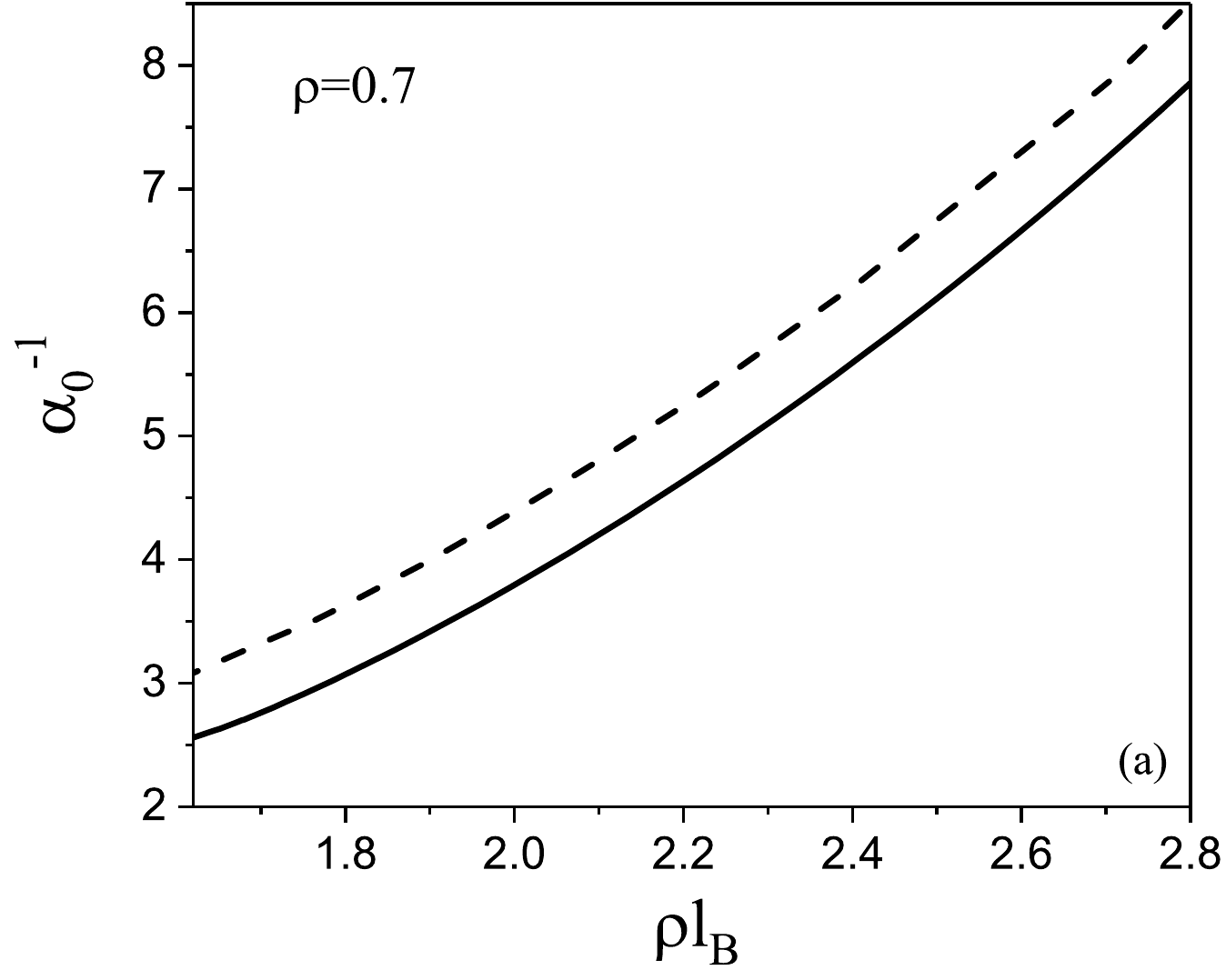}
	\includegraphics[scale=0.25]{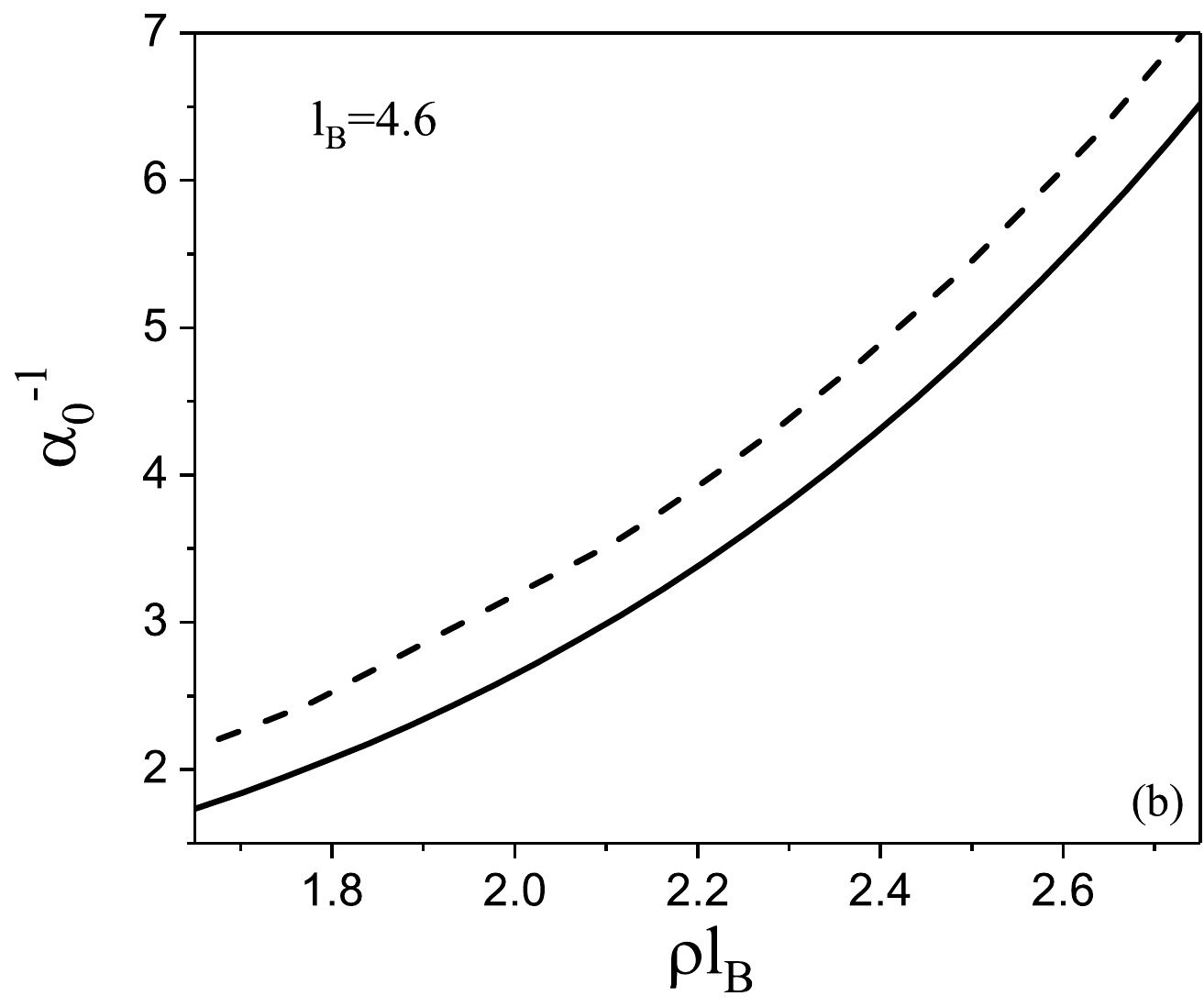}
	\caption{The correlation length $\alpha_0^{-1}$ of the charge--charge correlations  as a function of $\rho l_\text{B}$ obtained numerically as solution of  equations~(\ref{e1})--(\ref{cr2}) (solid line) and without the last term in (\ref{Ccc3}) (dashed line)  for the fixed dimensionless density of ions $\rho=0.7$ (a) and for the fixed Bjerrum length $l_{B}=4.6$~(b). }
	\label{fig:alpha0^-1_2}
\end{figure}

\begin{figure}[!t]
	\centering
	\includegraphics[scale=0.25]{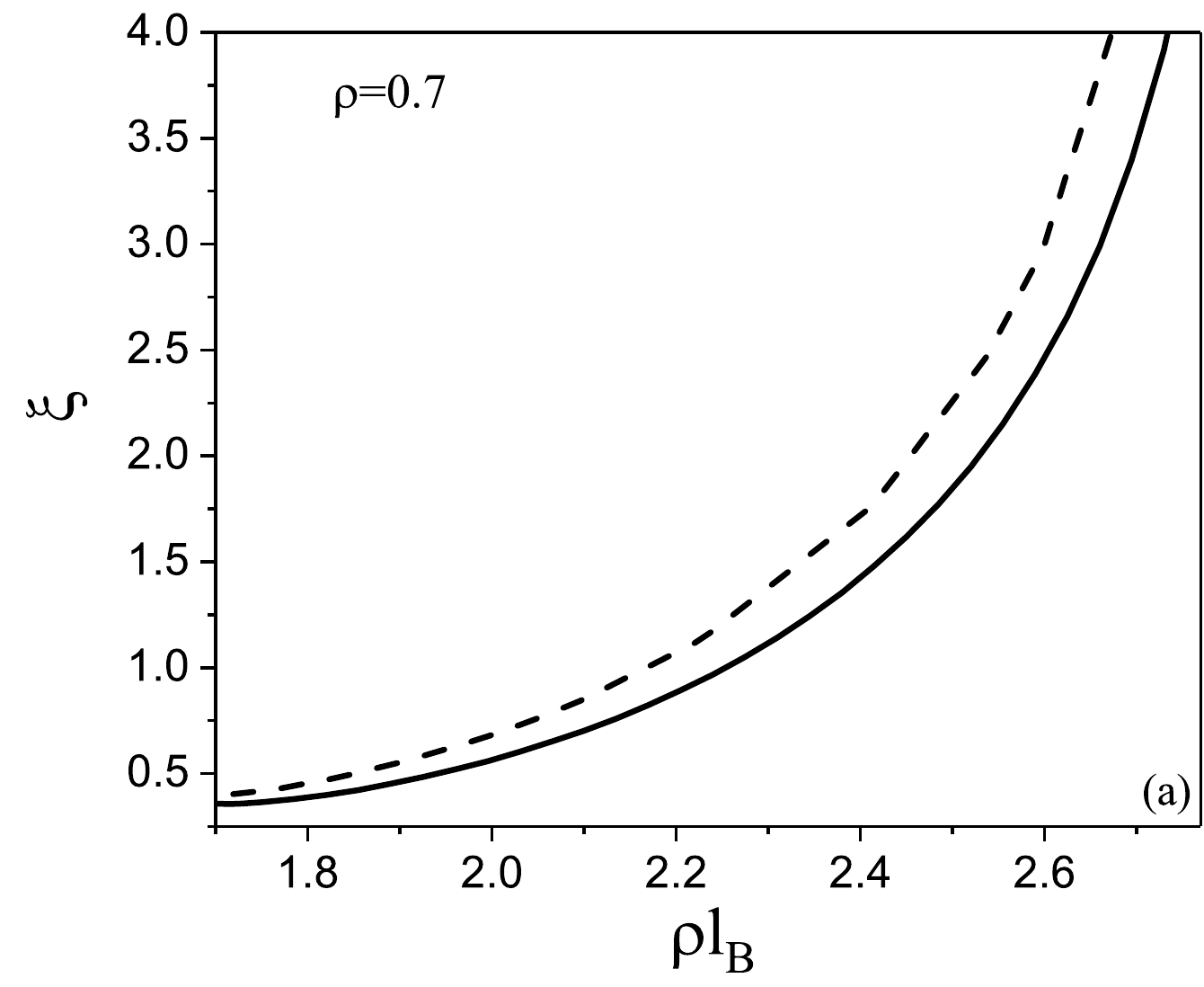}
	\includegraphics[scale=0.25]{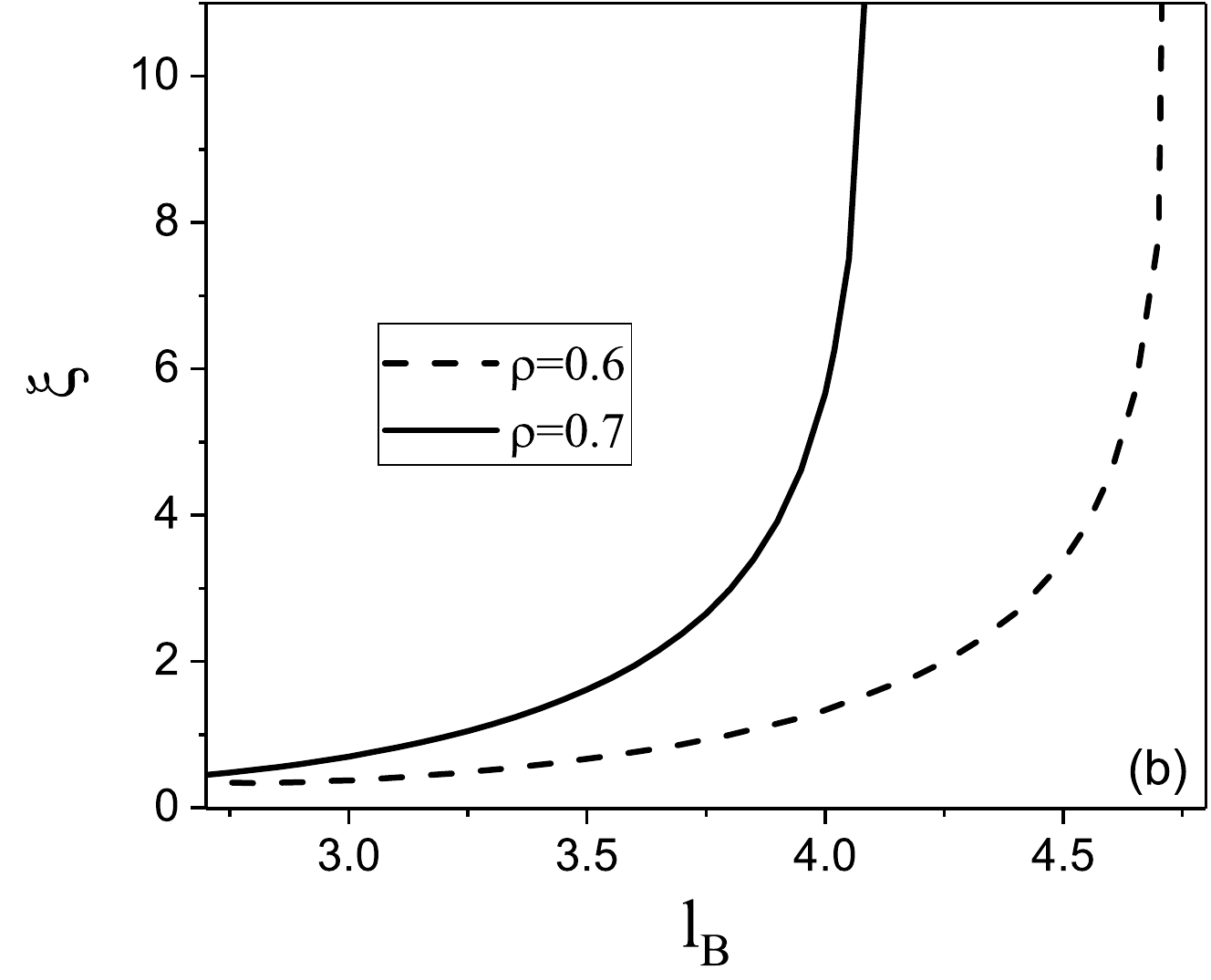}
	\caption{The correlation length  $\xi$ of the density--density correlations. Panel (a): obtained numerically as a solution of the equations~(\ref{e1})--(\ref{cr2}) (solid line) and without the last term in (\ref{Ccc3}) (dashed line)  for  the fixed dimensionless density of ions $\rho=0.7$. Panel (b):  with numerical solution of  equations~(\ref{e1})--(\ref{cr2})  for  the fixed dimensionless density of ions $\rho=0.6$ {(dashed line)} and $\rho=0.7$ {(solid line)}.}
	\label{fig:xi}
\end{figure}

\begin{figure}[!t]
	\centering
	\includegraphics[scale=0.25]{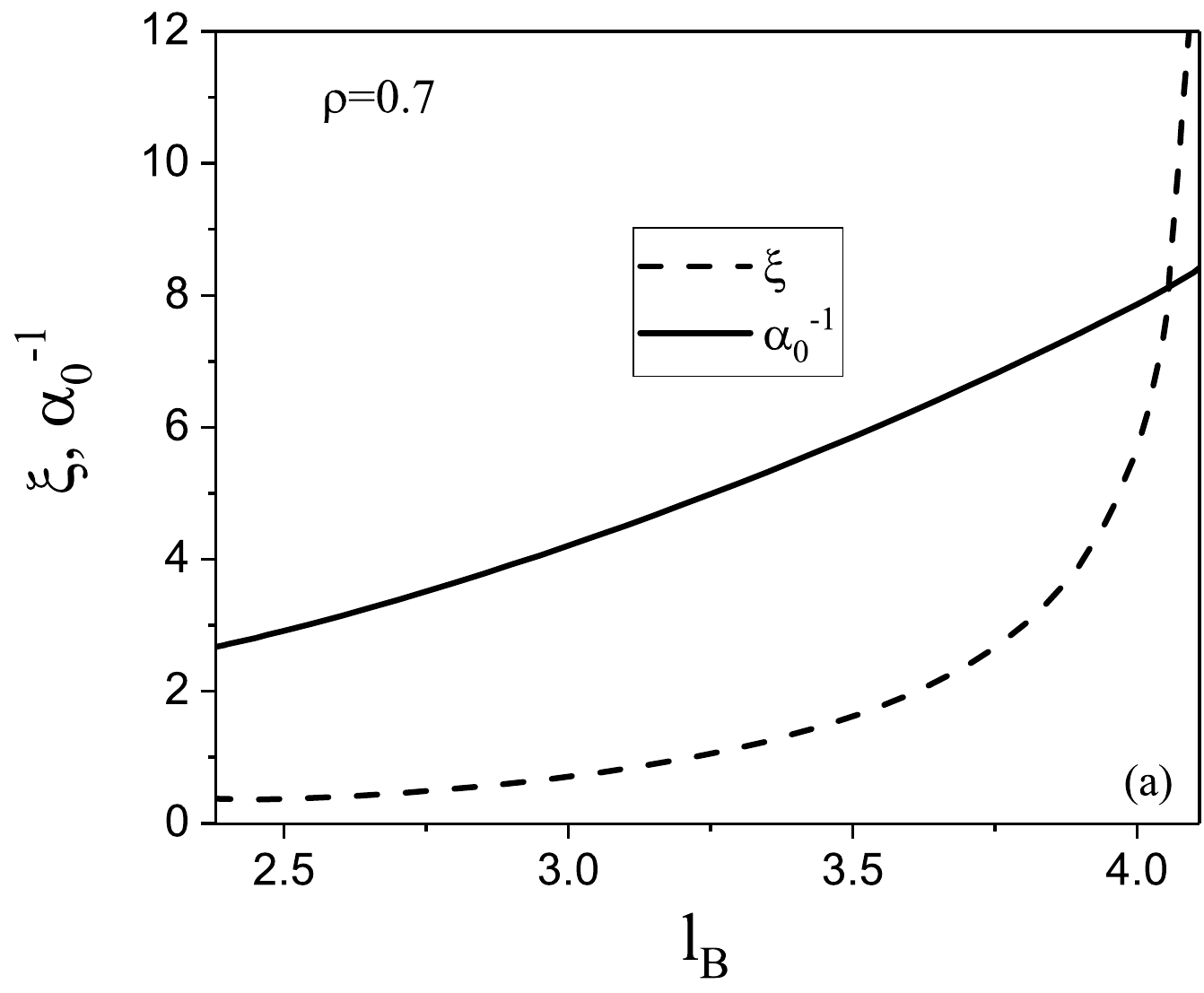}
	\includegraphics[scale=0.25]{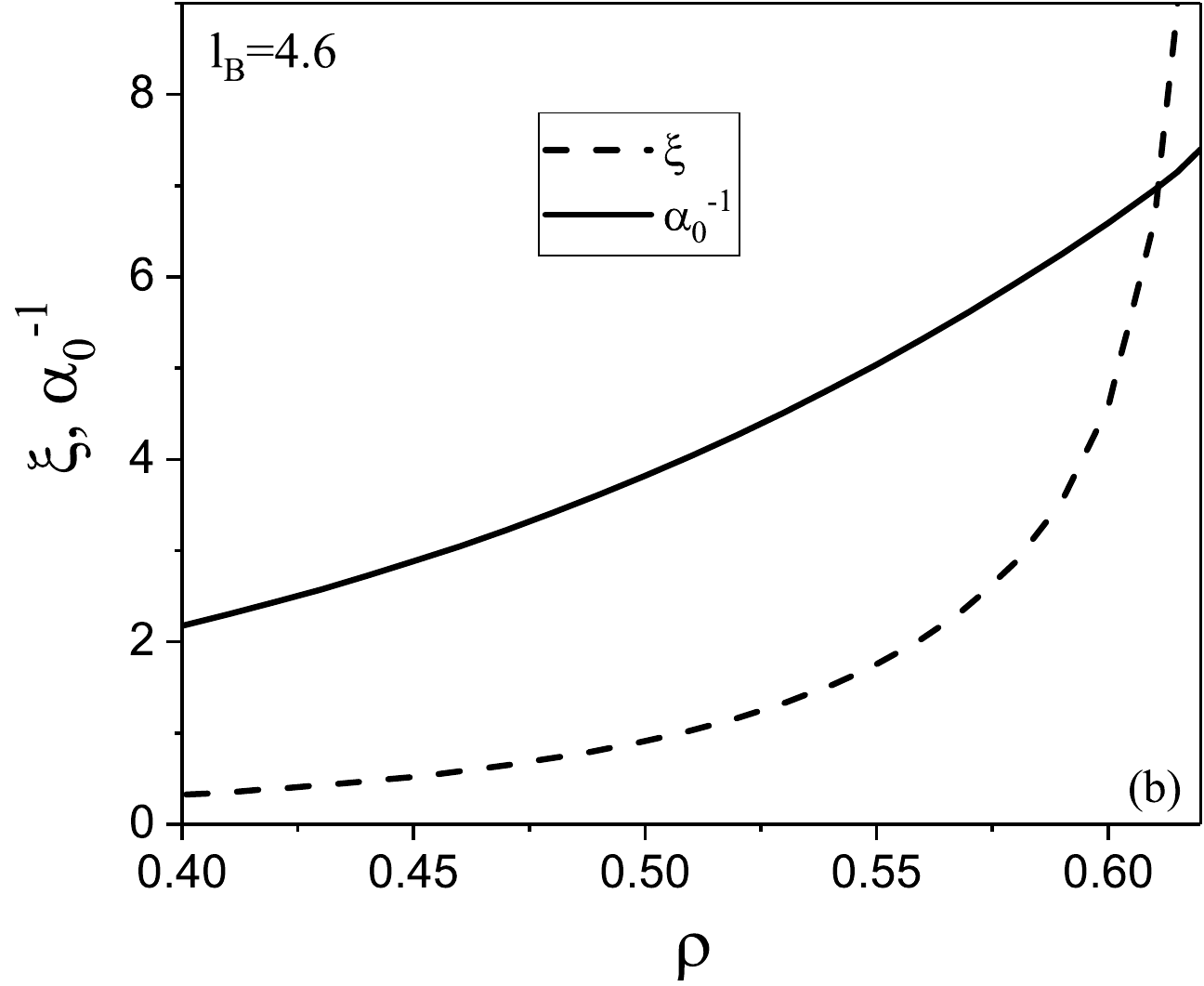}
	\caption{The correlation lengths  $\alpha_0^{-1}$  {(solid lines)} and $\xi$  {(dashed lines)} of the charge--charge and density--density correlations, respectively,   as functions of the Bjerrum length $l_\text{B}$ for  the fixed dimensionless density of ions $\rho=0.7$ (panel~a) and as functions of the dimensionless density of ions $\rho$ for the fixed Bjerrum length $l_\text{B}=4.6$ (panel~b).
	}
	\label{fig:xi_alpha0_lB}
\end{figure}

The correlation lengths of charge--charge  and density--density correlations,  $\alpha_0^{-1}$ and $\xi$, calculated using equations (\ref{e1})--(\ref{cr2}), are shown in figures~\ref{fig:alpha0^-1}--\ref{fig:xi_alpha0_lB}. The results are obtained for the ranges of dimensionless ion density $\rho$ and  Bjerrum length $l_\text{B}$ for which the condition (\ref{Caa}) is satisfied. 

In figure~\ref{fig:alpha0^-1}, the dependence of $\alpha_0^{-1}$ on the Bjerrum length $l_\text{B}$ for fixed high ion density $\rho=0.8$  is compared with our previous results \cite{ciach:23:1} obtained  without  the last term in  (\ref{Ccc3}) from the approximate analytical theory (dashed line) and with a numerical solution (dash-dotted line). It is seen that the lines $\alpha_0^{-1}(l_\text{B})$
obtained with and without the last term in  (\ref{Ccc3}) are almost parallel and
$\alpha_0^{-1}$ obtained from (\ref{e1})--(\ref{cr2}) is  closer to  the  results of the analytical theory converging with them for $l_\text{B}>2.4$.

The correlation length  $\alpha_0^{-1}$ calculated  numerically with and without the last term in (\ref{Ccc3}) is shown as a function of $\rho l_\text{B}$ for fixed $\rho=0.7$ [figure~\ref{fig:alpha0^-1_2},(a)] and for fixed $l_\text{B}=4.6$ [figure~\ref{fig:alpha0^-1_2},(b)]. For the fixed  density of ions,  $\alpha_0^{-1}$  depends on $\rho l_\text{B}$  almost linearly for $\rho l_\text{B}>1.8$. For the fixed Bjerrum length $l_\text{B}$, $\alpha_0^{-1}$ also
increases almost linearly with $\rho l_\text{B}$  but it occurs for larger $\rho l_\text{B}$. Both the solutions lead to a similar  behaviour of $\alpha_0^{-1}(\rho l_\text{B})$.

In figure~\ref{fig:xi} (a), we compare the dependences of the correlation length of density--density correlations  $\xi$ on $\rho l_\text{B}$ for the fixed density of ions $\rho=0.7$, obtained numerically with and without the last term  in (\ref{Ccc3}). Being almost the same in both cases for small $\rho l_\text{B}$, $\xi$ obtained from  the solution of  equations~(\ref{e1})--(\ref{cr2}) increases with increasing $\rho l_\text{B}$ more slowly  than  $\xi$ obtained when the last term in (\ref{Ccc3}) is neglected. Figure~\ref{fig:xi} (b) shows the correlation length  $\xi$ as a function of the Bjerrum length $l_\text{B}$ for two fixed ion densities $\rho=0.6$ and $\rho=0.7$. 

\begin{figure}[!t]
	\centering
	\includegraphics[scale=0.25]{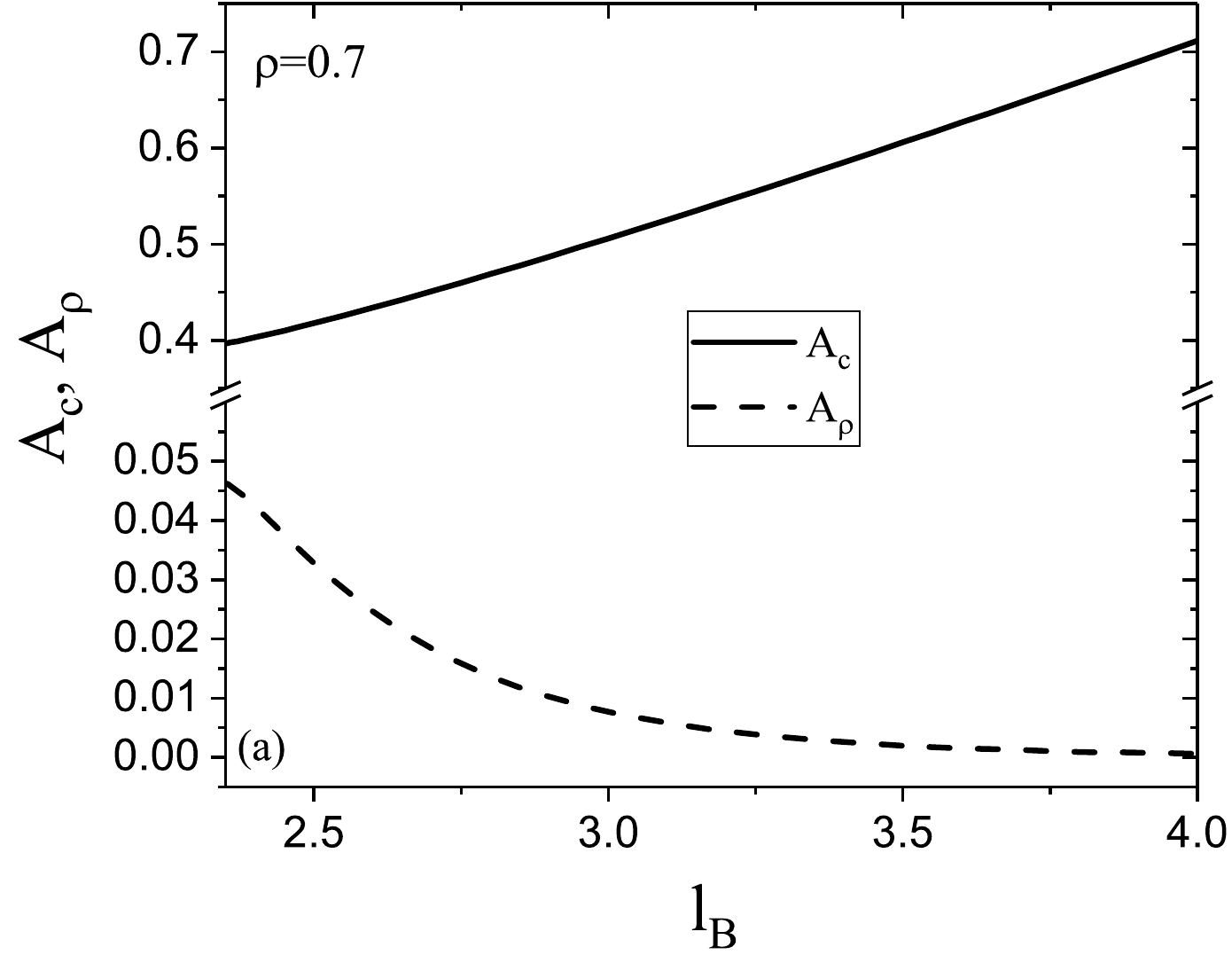}
	\includegraphics[scale=0.25]{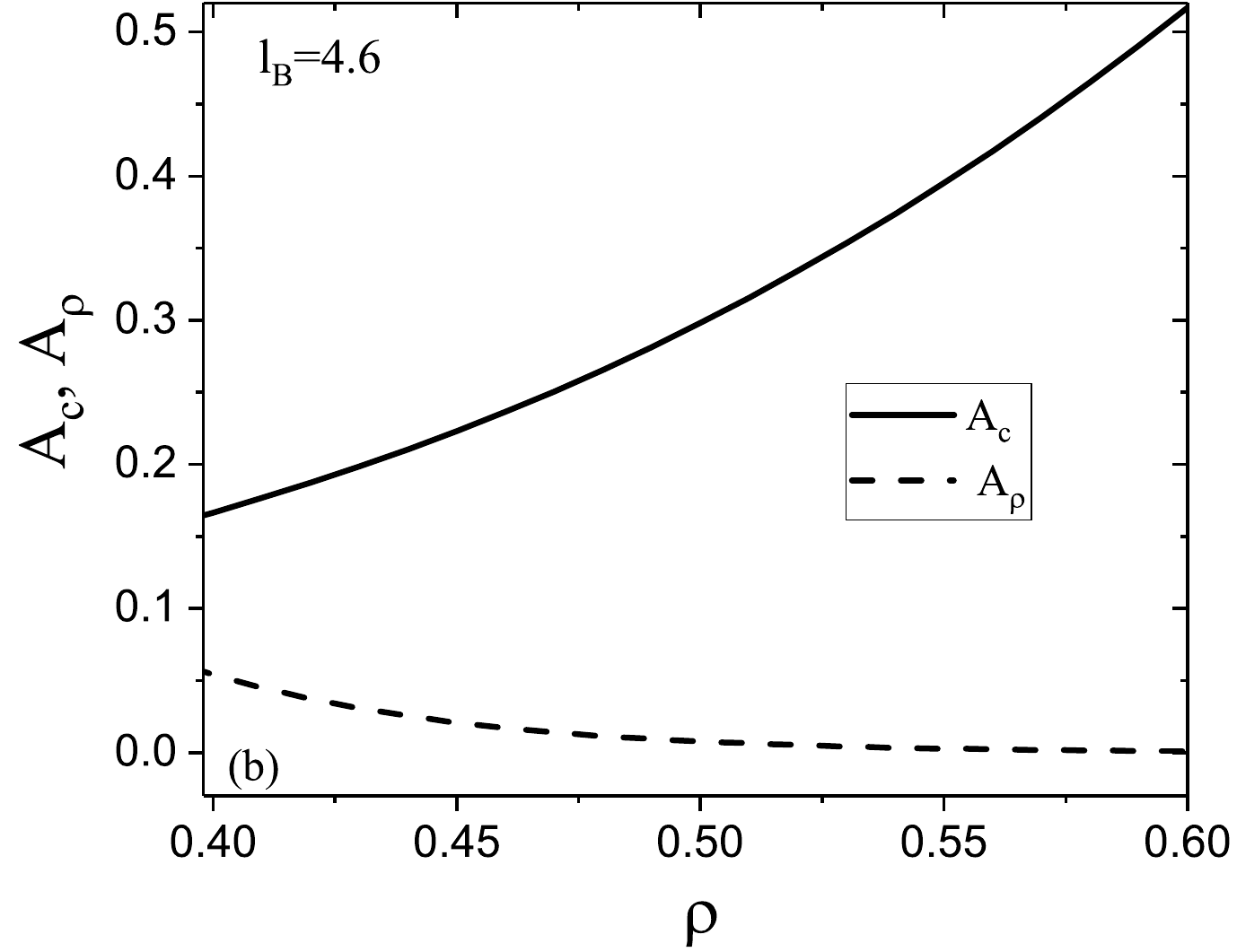}
	\caption{The amplitudes of the charge--charge and density--density correlation functions, $A_{cc}$ {(solid lines)} and $A_{\rho\rho}$  {(dashed lines)}, as functions of the Bjerrum length $l_\text{B}$ for  the fixed dimensionless density of ions $\rho=0.7$ (panel~a) and as functions of the dimensionless density of ions $\rho$ for the fixed Bjerrum length $l_\text{B}=4.6$ (panel~b).
	}
\label{fig:amplitudes}
\end{figure}

The correlation lengths  of the charge--charge and density--density correlations, $\alpha_0^{-1}$ and $\xi$,  are shown in figure~\ref{fig:xi_alpha0_lB} as functions of the Bjerrum length $l_\text{B}$ for the fixed ion density $\rho=0.7$ [panel (a)] and as  functions of $\rho$ for the fixed $l_\text{B}=4.6$ [panel (b)]. As in figure~\ref{fig:alpha0^-1_2}, we have almost a linear dependence of $\alpha_0^{-1}$  on $l_\text{B}$ (or on $\rho$). For  the fixed $\rho$, the correlation length  $\alpha_0^{-1}$ is larger than the correlation length $\xi$ for $l_\text{B}< 4$. A similar behaviour is observed for the  fixed $l_\text{B}$ for $\rho< 0.6$. It is worth noting that  
the crossover point separating the charge-dominated  and density-dominated regimes shifts towards a higher concentration as the dielectric constant decreases \cite{Yang2023,Jger2023,wang:24:0}. 
 The crossover point for the both cases shown in figure~\ref{fig:xi_alpha0_lB} is at $\rho l_\text{B}\approx 2.8$ corresponding to $a/\lambda_\text{D}\approx 6$. For the RPM, the crossover point at $a/\lambda_\text{D}\approx 3$ was obtained theoretically \cite{Cats2021} and from the simulations \cite{Yang2023}. 
 Such a rather large difference may be due to the fact that our theory does not take into account the packing of hard spheres for high densities.

In figure~\ref{fig:amplitudes}, the amplitudes of the charge--charge and density--density correlation functions, $A_{cc}$ and $A_{\rho\rho}$, calculated using equations (\ref{Ac1})--(\ref{Ar1}) are presented
for the same values of fixed $\rho$ and $l_\text{B}$  as in figure~\ref{fig:xi_alpha0_lB}. For both, the fixed $\rho$ and the fixed $l_\text{B}$, $A_c$  increases  and $A_\rho$ decreases  with an increase of $l_\text{B}$ and $\rho$, respectively. Moreover, $A_\rho\approx 0.001$  is  two orders of magnitude smaller than $A_c$  for $l_\text{B}\ge 4$ for fixed $\rho=0.7$ and for $\rho\ge 0.6$ for fixed $l_\text{B}=4.6$.

\section{Discussion}
\label{discussion}
In this section we discuss the controversial issue of the asymptotic decay of the charge-charge correlations at a more general level. The experiments~\cite{smith:16:0,lee2017} and recent simulations~\cite{hartel2024} suggest that $G_{cc}(r)$ is a sum of terms with oscillatory and monotonous decays, with the latter having long range and small amplitude. The decay lengths are determined by zeros of $\hat C_{cc}(k)$ extended to the complex $q$ plane.
In our mesoscopic theory, there is a pair of complex-conjugate poles or two imaginary poles of  $\hat G_{cc}(q)=1/\hat C_{cc}(q)$ when the  contribution to $C_{cc}$ associated with the correlation of the  simultaneous charge and density fluctuations,
\begin{equation}
\label{Jccr}
\beta J_{cc}(r)=-\langle \phi({\bf r}_1)\psi({\bf r}_1)\phi({\bf r}_2)\psi({\bf r}_2)\rangle/\rho^4\approx -G_{cc}(r)G_{\rho\rho}(r)/\rho^4,
\end{equation}
is neglected. 
In (\ref{Jccr}), $r=|{\bf r}_1-{\bf r}_2|$, and the approximate equality on the right-hand side holds in the Gaussian approximation.
 Thus,  when the only dependence of $\hat C_{cc}(k)$ on $k$ comes from the Coulomb potential, $G_{cc}(r)$   either consists of two terms decaying monotonously, or exhibits an oscillatory decay.
 The inclusion of the fluctuation contribution (\ref{Jccr}) to $C_{cc}(r)$  
 brings additional dependence of $\hat C_{cc}(k)$ on $k$, and it might lead to an additional pole of $\hat G_{cc}(q)$. This pole  in turn might lead to 
additional  monotonously decaying  term  in $G_{cc}(r)$ that exhibits an oscillatory decay for large density of ions. In any case, an extra $k$-dependent term in $\hat C_{cc}(k)$ is necessary for additional zero
 of $\hat C_{cc}(q)$ on the complex $q$-plane.
 
 We have shown, however, that in the mesoscopic theory at the level of the Gaussian approximation, 
the coupling of the charge and density fluctuations does not lead to a significant change of the correlation functions. In the Gaussian approximation, the contribution associated with the simultaneous charge and density fluctuations has the form  given by (\ref{Jcc}). 
In the mesoscopic theory, however, the packing of hard spheres for large densities that leads to an oscillatory decay of $G_{\rho\rho}(r)$ for small distances with the period about half the period of $G_{cc}(r)$ is not taken into account. We also neglected noncoulombic interactions and  fluctuation contributions to the higher-order correlation functions, assuming $\langle \phi({\bf r}_1)\psi({\bf r}_1) \phi({\bf r}_2)\psi({\bf r}_2)\rangle\approx G_{cc}(r)G_{\rho\rho}(r)$.  
It might be possible that our approximation for the $k$-dependent contribution to $\hat C_{cc}(k)$ other than the one following from the Coulomb interactions
is oversimplified. 

Before developing various extensions of our theory, it is instructive to consider an inverse problem. We assume that $G_{cc}(r)$ is a sum of the oscillatory and monotonous decays, and put a question:
what are the necessary conditions that should be satisfied by the  
fluctuation contribution  to $\hat C_{cc}(k)$ that comes from the 
charge-density coupling 
[equation(\ref{Jccr})]. 
Since the monotonous decay found in experiments~\cite{smith:16:0,lee2017} and simulations~\cite{hartel2024}  is long-range, the small $k$ behavior of  $\hat C_{cc}(k)$ is relevant. We postulate that the Taylor-expanded term following from the charge-density coupling has the form 
$\beta \hat J_{cc}(k)=c_0+c_2k^2+c_4 k^4 +O(k^6)$.
The expansion can be truncated for $c_4>0$, because we require $\hat C_{cc}(k)>0$ for $k>0$ in the single-phase region. To determine the decay lengths, we consider $k^2 \hat C_{cc}(k)$, and the Taylor expansion of (\ref{Ccc3}) valid for $k\ll 1$ takes the form
\[
k^2 \hat C_{cc}(k)=4\piup l_\text{B}\Bigg[1 +\Big(\frac{ -2\piup l_\text{B} +\rho_R^{-1}+c_0}{4\piup l_\text{B}}\Big) k^2 +\Big(\frac{\piup l_\text{B}/6 +c_2}{4\piup l_\text{B}}\Big) k^4+\Big(\frac{-\piup l_\text{B}/180 +c_4}{4\piup l_\text{B}}\Big) k^6\Bigg].
\]
{The above equation indicates that the Stillinger-Lovett conditions  \cite{stillinger:68:0} [$\hat{G}(k)=k^2/(4\piup l_\text{B})$ for $k\rightarrow 0 $] are held in the self-consistent Gaussian approximation considered in this work.}

We define
$P(x)= 1+b_1 x+ b_2 x^2 + b_3 x^3$ with 
\begin{equation}
\label{b}
b_1=\frac{ -2\piup l_\text{B} +\rho_R^{-1}+c_0}{4\piup l_\text{B}}, \quad b_2=\frac{\piup l_\text{B}/6 +c_2}{4\piup l_\text{B}}, \quad b_3=\frac{-\piup l_\text{B}/180 +c_4}{4\piup l_\text{B}}.
\end{equation}
In general, if $b_3>0$ then $P(x)$ has a real root $x_m<0$  corresponding to $k_m=\ri\sqrt{|x_m|}$. 
In addition, $P(x)$ has either two more real roots or a pair of complex-conjugate roots. Large correlation length means small $|x_m|$, which implies large  $b_1>0$. 

The oscillatory decay of $G_{cc}(r)$ is associated with the real part of the root of $ \hat C_{cc}(\alpha_1+\ri\alpha_0)=0$ that is $\alpha_1\approx 2$, and the Taylor expansion may be not valid for this region of $k$. We assume that the oscillatory decay of  $G_{cc}(r)$ for large $\rho l_\text{B}$ can be determined by the vicinity of the minimum of $ \hat C_{cc}(k)$, and is only weakly affected by the charge-density coupling. This assumption is based on the semi-quantitative agreement of such approximate theory with simulations of the RPM. In fact, the precise value of $k_0$ obtained in our theory with neglected $\beta J_{cc}$ is somewhat different from the simulation results.
We require that $\beta \hat J_{cc}(k)$ does not lead to a large shift of the minimum of  $ \hat C_{cc}(k)$, and
\[
\frac{\rd\hat C_{cc}(k)}{\rd k}=-\frac{4\piup l_\text{B}\sin(k)}{k^2}-\frac{8\piup l_\text{B}\cos(k)}{k^3}+2c_2 k+4c_4k^3=0
\]
holds for $k\sim k_0$. This is possible when $ c_2 +2c_4k_0^2\approx 0$
and we conclude that the minimum of $\hat C_{cc}(k)$ with and without the charge-density coupling can occur at similar $k$ when $c_2<0$.

The necessary conditions for the existence of the monotonous and oscillatory decays of $G_{cc}(r)$ are $b_1>0, b_2<0, b_3>0$. The large decay length of the monotonous decay can occur when $b_1\gg 1$, because $P(x)=0$ for $x<0$ with $|x|\ll 1$ only for a very large slope of $P(x)$ at $x=0$ when $P(0)=1$.

For $\rho_R$ that is not very small,  $b_1\gg 1$ is possible for large $c_0$, which in turn implies $b_1\approx c_0/(4\piup l_\text{B})$. Note that  when $b_1\gg b_2,b_3$ the decay length is approximately $\sqrt{b_1}$, because $k^2 \hat C_{cc}(k)\approx 4\piup l_\text{B}[1+b_1 k^2+...]$ for $k\ll 1$. If we require that the decay length should be $\sqrt{b_1}\propto \rho l_\text{B}$ as in the experiments, we obtain $c_0=\beta \hat J_{cc}(0)\propto \rho^2 l_\text{B}^3$.

The above discussion shows that the charge-charge correlation function could contain a monotonously decaying contribution for large $\rho l_\text{B}$ if the necessary 
condition $\beta \hat J_{cc}(0)\gg 4\piup l_\text{B}$ 
was satisfied.
The above and the scaling for the correlation length of the asymptotic monotonous decay of $G_{cc}(r)$ would be simultaneously satisfied if
\begin{equation}
\label{necessary}
\beta \hat J_{cc}(0)=A \rho^2 l_\text{B}^3\gg 4\piup l_\text{B},
\end{equation}   
where $A>0$ is a proportionality constant that must satisfy $A\gg 4\piup/(\rho l_\text{B})^2$.

In the Gaussian approximation $\beta\hat J_{cc}(0)$
can be large and positive when the integral of $G_{cc}(r)G_{\rho\rho}(r)$ is negative. In this approximation
 and with ordering of hard spheres and noncoulombic interactions both neglected, the above necessary conditions cannot be satisfied by the correlation functions that have physical meaning.   
The question whether in some
extensions of our theory
the necessary condition (\ref{necessary}) as well as  $ b_2<0, b_3>0$ (see (\ref{b})) can be satisfied remains open.   
\section{Conclusions}
We have shown that the addition of the previously neglected fluctuation contribution to the inverse charge-charge correlation function does not lead to a significant change of the results, as long as we restrict ourselves to the Gaussian approximation in our mesoscopic theory. 
The presence of the term with the correlation between simultaneous charge and density fluctuations [the last term in (\ref{Ccc3})]  makes the problem of self-consistent solution of the equations for the charge-charge and density-density correlations much more complex. To make the problem tractable, we limited ourselves to a rather small ranges of the density and the Bjerrum length, where the simplified equations are valid. Our results confirm that within the Gaussian approximation, the previous results concerning the correlation functions remain valid on the semi-quantitative level. 
 
 We also derived necessary conditions for the existence of a monotonous decay  in addition to the oscillatory decay of $G_{cc}(r)$ for large $\rho l_\text{B}$. We have shown that these conditions cannot be satisfied within our theory at the level of the Gaussian approximation.  
 
 Beyond the Gaussian approximation, additional terms are expected on the right-hand side in (\ref{Jccr}), and extension of the present theory is necessary to verify whether these terms can lead to a change of $\beta \hat J_{cc}(0)$ that is big enough to allow for fulfilling the necessary conditions for the asymptotic monotonous decay of $G_{cc}(r)$ for large $\rho l_\text{B}$.
  
 Note that if only the charge changes in the mesoscopic region and the number of ions remains constant, then an ion is replaced by an ion with the opposite charge.  If only a number of ions changes and the charge remains zero, then a pair of oppositely charged ions enters or leaves the considered region. Two ions are engaged in these events.
 Simultaneous fluctuations of the local charge and density,   $\phi({\bf r})\psi({\bf r})$  are associated with a single ion entering or leaving the considered mesoscopic region. Thus, the simultaneous charge and density fluctuations can dominate, even though at the mathematical level they are represented by the four-point function. A more careful study of $\langle \phi({\bf r}_1)\psi({\bf r}_1)\phi({\bf r}_2)\psi({\bf r}_2)\rangle$, going beyond the Gaussian approximation can shed more light on the asymptotic decay of $G_{cc}(r)$. We shall focus on this question in our future studies.  
 
 { Let us finally comment that the mesoscopic theory can be applied to the studies of concentrated 
electrolytes near a charged electrode~\cite{ciach:18:1,ciach:25:0} or between two electrodes~\cite{ciach:2023:2}. 
 In addition to the charge distribution, it is possible to calculate a differential capacitance of the double 
 layer or of a planar capacitor, as well as the disjoining pressure~\cite{ciach:2023:2}. 
 The effects of the oscillatory decay of the charge density with the distance from the electrode on the 
 capacitance of the double layer have been determined so far only for very small voltage in~\cite{ciach:25:0}. 
A more detailed analysis in the future studies can shed light on the properties of the capacitance, including the 
 anomalies found in simulation and theoretical studies~\cite{boda:99:0,DiCaprio2006}.}   

\section{Acknowledgement}  
The article is dedicated to the 100th anniversary of Professor Ihor Yukhnovskii --- the prominent  Ukrainian scholar and  inspiring academic teacher. OP acknowledges with deep gratitude his influence on her throughout the years in the role of teacher and mentor.

\bibliographystyle{cmpj} 
\bibliography{bibliography_CMP_2024}

%
%

\ukrainianpart

\title{Взаємний вплив кореляцій густини заряду і числа частинок в іонних рідинах і концентрованих електролітах}
\author{О. Пацаган\refaddr{label1}, А. Цях\refaddr{label2}}
\addresses{
\addr{label1}Iнститут фiзики конденсованих систем Нацiональної академiї наук України
79011, м. Львiв, вул.~Свєнцiцького, 1, Україна
\addr{label2} Інститут фізичної хімії Польської академії наук, 01-224 Варшава, Польща
}

\makeukrtitle

\begin{abstract}
\tolerance=3000%
Кореляційні функції в концентрованих іонних системах вивчаються в рамках мезоскопічної теорії на рівні гауссового наближення. Флуктуаційний внесок в обернену кореляційну функцію ``заряд-заряд'', яким раніше нехтувалося, враховується для перевірки точності попередніх результатів. Ми розраховуємо кореляційні довжини та амплітуди і показуємо, що врахування  флуктуаційного внеску не призводить до суттєвих змін результатів. Також ми виводимо необхідні умови для існування як осциляційного, так і монотонного загасання кореляцій ``заряд-заряд'', які мають  виконуватися для некулонівських внесків в обернені кореляційні функції ``заряд-заряд''. На рівні гауссового наближення ці умови не виконуються. Необхідним є розширення теорії за рамки гауссового наближення для того, щоб перевірити, чи  асимптотичне загасання кореляцій ``заряд-заряд'' є монотонним чи осциляційним, як це було отримано, відповідно, в експериментах з вимірювання поверхневих сил чи в експериментах методом малокутового рентгенівського розсіяння.

\keywords кореляційна функція ``заряд-заряд'', кореляційні довжини, концентровані електроліти, мезоскопічна теорія
\end{abstract}

\end{document}